\documentclass{article}
\usepackage{authblk}

\usepackage[english]{babel}
\usepackage{ulem}
\usepackage[letterpaper,top=2cm,bottom=2cm,left=3cm,right=3cm,marginparwidth=1.75cm]{geometry}

\usepackage{assumptionsofphysics}
\usepackage[dvipsnames]{xcolor}
\usepackage{tikz-cd}
\usepackage[colorlinks=true, allcolors=blue]{hyperref}
\usepackage{pgfplots}

\def\>{\rangle}
\def\<{\langle}

\title{Classical mechanics as the high-entropy limit \\ of quantum mechanics}
\author[1]{Gabriele Carcassi}
\author[2]{Manuele Landini}
\author[1]{Christine A. Aidala\thanks{Corresponding author.~\href{mailto:caidala@umich.edu}{caidala@umich.edu}}}

\affil[1]{Physics Department, University of Michigan, Ann Arbor, Michigan 48109, USA}

\affil[2]{Institut f\"ur Experimentalphysik und Zentrum f\"ur Quantenphysik, \protect\\ Universit\"at Innsbruck, Innsbruck, Austria}

\begin{document}
\maketitle

\begin{abstract}
We show that classical mechanics can be recovered as the high-entropy limit of quantum mechanics. That is, the high entropy masks quantum effects, and mixed states of high enough entropy can be approximated with classical distributions. The mathematical limit $\hbar \to 0$ can be reinterpreted as setting the zero entropy of pure states to $-\infty$, in the same way that non-relativistic mechanics can be recovered mathematically with $c \to \infty$. Physically, these limits are more appropriately defined as $S \gg 0$ and $v \ll c$. Both limits can then be understood as approximations independently of what circumstances allow those approximations to be valid. Consequently, the limit presented is independent of possible underlying mechanisms and of what interpretation is chosen for both quantum states and entropy. 
\end{abstract}

\section{Introduction}
\label{sec:intro}

The goal of this paper, summarized in Fig.~\ref{fig_quadrants}, is to show that classical mechanics is the high-entropy limit of quantum mechanics, much in the same way that it is the low-speed limit of relativistic mechanics.

\begin{figure}[h]
	\centering
\begin{tikzpicture}
    % Box dimensions
    \def\boxwidth{4}
    \def\boxheight{2}
    
    % Draw four boxes
    \draw (0,0) rectangle (\boxwidth,\boxheight);
    \draw (-\boxwidth,0) rectangle (0,\boxheight);
    \draw (0,-\boxheight) rectangle (\boxwidth,0);
    \draw (-\boxwidth,-\boxheight) rectangle (0,0);
    
    % Label Boxes
    \node[align=center] at (\boxwidth/2,\boxheight/2) {Relativistic \\ Mechanics};
    \node[align=center] at (-\boxwidth/2,\boxheight/2) {Classical \\ Mechanics};
    \node[align=center] at (-\boxwidth/2,-\boxheight/2) {Quantum \\ Mechanics};
    \node[align=center] at (\boxwidth/2,-\boxheight/2) {Quantum \\ Field Theory};

    % Axes
    \draw[->] (-\boxwidth,\boxheight+0.5) -- (\boxwidth,\boxheight+0.5) node[above,midway] {Speed};
    \draw[->] (-\boxwidth-0.5,-\boxheight) -- (-\boxwidth-0.5,\boxheight) node[above,midway,rotate=90] {Entropy};
    
\end{tikzpicture}
	\caption {The amended four-quadrant picture often used to compare the theories. The distinction between classical and quantum physics is not size, but entropy. Small systems in states of high entropy (e.g.~a few molecules at high temperature) can be described by classical mechanics; large systems in states of low entropy (e.g.~systems entangled over long distances) cannot be described by classical mechanics. } \label{fig_quadrants}
\end{figure}

Physically, as we take states at higher and higher entropy, the quantum features become less and less pronounced to the point that classical mechanics becomes a good approximation. This limit is therefore independent of what interpretation one may give to either quantum mechanics or entropy, and from the reason, the mechanism, that makes a high-entropy description suitable. Quite literally, quantum mechanics is needed exactly when we need more precise (i.e.~low-entropy) descriptions. This brings the classical limit in line with the non-relativistic limit, where the low-speed limit is seen as unrelated to discussions about the nature of space-time and no one tries to find a reason, a mechanism, as to why speeds become low in certain regimes.

Mathematically, it is common to see both limits as group contractions. This too can be given a physically meaningful interpretation. The low-speed limit is mathematically recovered by taking the highest possible speed and setting it to infinity. This way, all finite velocities are small compared to the highest possible speed. This corresponds to taking $c \to \infty$. Conversely, the high-entropy limit is mathematically recovered by taking the lowest possible entropy, zero in quantum mechanics, to minus infinity. Note that entropy is the logarithm for the count of states, which is quantified in units of $\hbar$. Therefore, taking $\hbar \to 0$ is equivalent to setting the entropy for pure states to minus infinity. This way, all finite entropies are large compared to the lowest possible entropy. In a nutshell, this is what we will show in section~\ref{sec:high_entropy_limit}. Since physically it does not make sense to take limits of physical constants, $c \to \infty$ should be understood as $v \ll c$, $\hbar \to 0$ should be understood as $S \gg 0$.

The key insight is that \textbf{high entropy is the one and only requirement}. We do not require a particular interpretation of quantum mechanics, of entropy or a particular mechanism responsible for the high entropy. Any will do.\footnote{The only requirement for entropy is that it matches the experimentally measured one and that its maximization corresponds to the correct experimental quantities. In this vein, the Shannon/Gibbs entropy of classical mechanics and the von Neumann entropy of quantum mechanics are simply the tools we use to make predictions within the respective theories. The Boltzmann entropy coincides with the Shannon/Gibbs entropy for a uniform distribution. }

The paper is structured as follows. In the second section we give intuitive insights as to why the limit makes sense on physical grounds. We will argue that to produce quantum effects one does effectively prepare, in one way or another, states at low entropy. We will study the relationship between uncertainty and entropy in both classical and quantum mechanics, and see how closely these coincide even with an uncertainty of just a few $\hbar$. Lastly, we will see how quantum states at high entropy become closer to each other, and how uncertainty on quantum features can become equivalent to uncertainty on classical features. This aliasing is what makes the limit possible. In the third section we will give an example of how traditional limits can be reinterpreted as high-entropy limits. In the fourth section we will show how the high-entropy limit, mathematically, is like taking $\hbar \to 0$. We will study entropy-increasing transformations in classical mechanics and see that these are stretching operations over phase space. We then find quantum analogues and note that their effect is to rescale the commutator between position and momentum by a constant. Mathematically, this can be understood as rescaling $\hbar$. Finally, to give a more physically meaningful insight to the limit, we look at how the Wigner function changes as the phase space is stretched and the entropy increases.

\section{High entropy and classical states}

Since we aim to provide an approach to the classical limit that makes intuitive sense to all those that routinely work with quantum systems, we will start with a few qualitative considerations that hint at the connection between classical mechanics and high entropy.

\subsection{Producing quantum states}

One of the experimental challenges in quantum mechanics is producing states that are ``quantum enough'' to exhibit quantum properties. What we want to show is that all these problems can ultimately be understood as reducing the entropy of the initial state.

Coherence is probably one of the direct and most important properties. It has been shown that coherence can be maintained over long distances and among a large number of constituents, meaning that quantum systems are not necessarily small or made of few components. However, it is also established how coherence can be quickly lost through interaction with the environment, through decoherence~\cite{Zurek_RevModPhys.75.715, SCHLOSSHAUER20191}. Since decoherence increases the entropy of the system, it represents one mechanism to reach the high-entropy limit. This is in line with our result.

In experimental practice, many quantum effects (e.g.~superconductivity, topological insulators, quantum Hall effect, ...) are harder or impossible to achieve at high temperature. The thermal noise can break the coherence of the system. Note that entropy is a monotonic function of temperature, meaning that decreasing temperature means decreasing entropy of the system. This is in line with our result.

Some quantum effects can be replicated at higher temperature given a high pressure. This is the case, for example, in some superconductive materials~\cite{Drozdov:2015}. High pressure corresponds to low entropy, which is in line with our result. Along the same lines, white dwarfs and neutron stars exhibit quantum effects at high pressure, despite high temperatures.

To produce Bose-Einstein condensates, one needs a high density in phase space, which means both high spatial density (to decrease the position spread) and low temperature (to decrease the momentum spread). The phase space density (PSD) is directly related to entropy and dimensionality, as exemplified by the famous Sackur-Tetrode equation for the entropy of the ideal gas. This, again, is in line with our result.

For a trapped condensate, the PSD at the center and entropy $S$ are related by $\text{PSD}=e^{5/2+\gamma-S/N}$, where $N$ is the number of particles and $\gamma$ is the virial coefficient for the trapping potential~\cite{PhysRevLett.78.990}. This has been used experimentally to condense at constant entropy, by changing the shape of the potential~\cite{PhysRevLett.81.2194}. Note that, however, the system is inhomogeneous, so the condensation happens because entropy is transferred from one part of the system to the other.

For thermodynamic systems in equilibrium, entropy can be calculated directly based on occupation of the quantum states of the system. For high temperatures, the occupations are all small and classical statistical mechanics applies. Quantum effects related to indistinguishability are found whenever occupations approach unity. The entropy per particle drops in the same limit, again in line with our result. Note, however, that while infinite entropy corresponds to infinite temperature, the reverse is not necessarily true. In systems with finitely many possible states we can find the stationary point of $S$ as a function of $E$ for which $1/T=\partial S/\partial E$ is zero. This stationary point represents the maximum entropy reachable by the system, and therefore there is no high-entropy limit. This is exactly the case where temperature can have negative values. Given that the entropy is a concave function of the energy, systems that admit a classical description are exactly those for which no such stationary point exists and temperature cannot be negative or infinite.%\footnote{{\color{blue}We can define temperature, based on the equation $1/T=\partial S/\partial E$. In systems with a finite number of degrees of freedom, we can find the stationary point of $S$ as a function of $E$. This is because both the minimum and maximum energy states will have $S=0$ or minimal. The maximum entropy state should correspond to infinite temperature. As a consequence, any microstate becomes equiprobable and the entropy takes the max value $S=k_B\log n_{dof}$} I reworked this point to again show how finitely many possibility does not give us a classical limit. }

\subsection{Uncertainty from entropy}

Another qualitative argument that shows the link between high entropy and classical mechanics comes from the relationship between entropy and uncertainty for a single degree of freedom.

In both classical and quantum mechanics, Gaussian states maximize entropy for a single independent degree of freedom at a fixed uncertainty or, equivalently, minimize uncertainty at fixed entropy. This means that, if we fix the entropy $S$, we will have an uncertainty relationship~\cite{Carcassi:2022bpm}
\begin{equation}
    \sigma_x \sigma_p \geq \Sigma(S).
\end{equation}
where $\sigma_x$ and $\sigma_p$ are the standard deviations of position and momentum and $\Sigma$ is the determinant of the covariance matrix. The specific value of the uncertainty $\Sigma$ will depend on the entropy and on whether we are using classical or quantum mechanics, though the relationship will always be saturated by Gaussian states with no correlation between $x$ and $p$.

In classical mechanics, the relationship between entropy and uncertainty for Gaussian states is~\cite{Cover_Thomas_2006,Pathria_Beale_2022}
\begin{equation}
    S_C(\Sigma) = \ln \left(2 \pi e \frac{\Sigma}{h}\right) = \ln \left(\frac{\Sigma}{\hbar}\right) + 1.
\end{equation}
In quantum mechanics it is~\cite{weedbrook2012gaussian}
\begin{equation}
S_Q(\Sigma) = \left( \frac{\Sigma}{\hbar} + \frac{1}{2} \right) \ln \left( \frac{\Sigma}{\hbar} + \frac{1}{2} \right) - \left( \frac{\Sigma}{\hbar} - \frac{1}{2} \right) \ln \left( \frac{\Sigma}{\hbar} - \frac{1}{2} \right).
\end{equation}

\begin{figure}
    \centering
\begin{tikzpicture}
\begin{axis}[
height=7cm,
width=\linewidth*0.8,
grid=both,
grid style={line width=.1pt, draw=gray!25},
axis lines=middle,
xlabel = \(\Sigma\),
ylabel = \(S\),
legend style={at={(0.95,0.65)},anchor=east},
]
\addplot[thick,blue,samples=150,domain=0.1:8.5] {ln(x)+1};
\addlegendentry{\(\text{classical}\)}
\addplot[thick,BrickRed,densely dotted,samples=150,domain=0.5:8.5] {(x+1/2)*ln(x+1/2)-(x-1/2)*ln(x-1/2)};
\addlegendentry{\(\text{quantum}\)}
\addplot[thick,ForestGreen,dashed,samples=150,domain=0.5:8.5] {(ln(x)+1) -((x+1/2)*ln(x+1/2)-(x-1/2)*ln(x-1/2))};
\addlegendentry{\(\text{difference}\)}
\end{axis}
\end{tikzpicture}
    \caption{Entropy $S$ in nats for a Gaussian state as a function of uncertainty $\Sigma$, measured in units of $\hbar$. In solid blue, the classical case $S = \ln(\Sigma) + 1$. In dotted red, the quantum case $S = \left( \Sigma + \frac{1}{2} \right) \ln \left( \Sigma + \frac{1}{2} \right) - \left( \Sigma - \frac{1}{2} \right) \ln \left( \Sigma - \frac{1}{2} \right)$. In dashed green the difference between the two.}
    \label{fig:uncertainty}
\end{figure}

In Fig.~\ref{fig:uncertainty} we can see that the classical and quantum cases are very close even when the uncertainty is just a few units of $\hbar$. Things diverge at about two units of $\hbar$: in quantum mechanics the entropy decreases faster and reaches zero at $\hbar/2$, the bound for the Heisenberg uncertainty principle; classical mechanics reaches zero entropy for $\hbar/e$, and then continues in the negative region.

The convergent prediction at high entropy together with the divergent prediction at low entropy, together with negative values for the classical case, reinforces the idea that quantum mechanics is required at low entropy much like relativity is required at high speeds.

\subsection{Entropic aliasing}

Note that, in our approach, the mechanism that increases the entropy is irrelevant: any will do. The intrinsic geometry of quantum mixed states makes uncertainty over classical or quantum properties look the same. It is this aliasing of different preparations that drives the classical limit, as we can see even in a standard two-slit experiment setup. 

In the simplest case, we can imagine the particle either passing through the left or right slit, which corresponds to the two pure states $|L\>$ and $|R\>$. These identify a two-state system, a qubit. We can now imagine equal superpositions of the two states, $\frac{1}{\sqrt{2}}|L\> + \frac{1}{\sqrt{2}} e^{-\imath \theta} |R\>$ where $\theta$ is the difference in phase between the two components. If $\theta$ is zero, we have the state $|+\> = \frac{1}{\sqrt{2}}|L\> + \frac{1}{\sqrt{2}} |R\>$ and the resulting interference pattern will have a peak in the middle of the screen. If $\theta$ is $\pi$, we have the state $|-\> = \frac{1}{\sqrt{2}}|L\> + e^{-\imath \pi}\frac{1}{\sqrt{2}} |R\>$ and the resulting interference pattern will have a valley in the middle of the screen.

As expected, $|+\>$ cannot be understood as a probability distribution, as a mixture, over $|L\>$ and $|R\>$. The interference pattern, the quantum feature, cannot be understood as a classical distribution over a path, a classical feature. However, an equal mixture of $|+\>$ and $|-\>$ is the maximally mixed state, which is equal to an equal mixture of $|L\>$ and $|R\>$. That is, the case where we randomize the phase is indistinguishable from the case where we randomize the path. In other words, uncertainty over a quantum feature behaves like uncertainty over a classical feature, regardless of the source of the uncertainty.

In the Bloch ball, states at equal entropy form a series of concentric spheres. As we increase the entropy, the points of these spheres become closer and closer, and therefore the error in using a mixture of $|L\>$ and $|R\>$ instead of the actual state becomes smaller. The same happens for the space of mixtures of any quantum system.

Mathematically, the trace distance $T(\rho,\sigma)$ between two mixed states $\rho$ and $\sigma$ is contractive under a completely positive trace-preserving (CPTP) map $M$. That is, $T(M(\rho),M(\sigma)) \leq T(\rho,\sigma)$~\cite{nielsen2010quantum}. In the finite-dimensional case, we can have a CPTP map that increases the entropy of all states except that of the maximally mixed state, which is a fixed point.\footnote{This is a particular instance of mixing/relaxing ergodic quantum channels where the fixed point is the maximally mixed state~\cite{Burgarth2012ErgodicAM}.}  Under such a map, all pure states are mapped to mixed states, orthogonal states are mapped to non-orthogonal states and the map is strictly contractive. Moreover, even if observables do not commute in general, they commute for the maximally mixed state $\rho = I/n$ in the following sense:
\begin{equation}
	\<AB\>_{\rho} = \tr(AB\rho) = \tr(ABI/n) = \tr(AB)/n = \tr(BA)/n = \tr(BAI/n)= \tr(BA\rho) = \<BA\>_{\rho}
\end{equation}
Observables for a state close enough to the maximally mixed state, then, will ``almost commute'' in the sense that the difference in the expectation of different permutations of their product will be small.

In section~\ref{sec:high_entropy_limit}, when we construct the high-entropy limit, the map that results will follow this intuition in the infinite-dimensional case, with the only difference that there is no maximally mixed state that does not change.\footnote{In the infinite-dimensional case, one could have infinite-dimensional subspaces that remain orthogonal while increasing the entropy of all its elements. This corresponds, for example, to the case where there are conserved quantities. Note that in the map we construct in section 4, the expectation of all polynomials of position and momentum changes, and therefore no function of position and momentum is a conserved quantity. }  If we call $\mathcal{G}$ the space of all mixtures of Gaussian states, the states with a non-negative Wigner function, given an arbitrary mixed state $\rho$, we will find a state $\sigma_{\rho} \in \mathcal{G}$ that is closest to $\rho$. As we apply our entropy-increasing map, the entropy of $\rho$ increases and $\sigma_{\rho}$ will get closer to $\rho$. In this sense the states will ``look classical.''

\section{Reinterpreting traditional approaches}

One of the advantages of our approach is that it is compatible with more traditional ones. Let us see how this works in two cases: the failure of classical statistical mechanics to predict black-body radiation and the recovery of classical statistical mechanics for thermal equilibrium.

\subsection{Black-body radiation}

The failure of classical statistical mechanics to predict the black-body radiation spectrum was one of the key drivers for the development of quantum theory. The spectrum predicted by classical mechanics is given by the Rayleigh–Jeans law,
\begin{equation}
    \frac{2 \nu^2 k_\text{B} T}{c^2} = \frac{2 \nu^2}{c^2 \beta},
\end{equation}
where $\beta = \frac{1}{k_\text{B} T}$. The one predicted by quantum mechanics is given by Planck's law,
\begin{equation}
    \frac{ 2 h \nu^3}{c^2} \frac 1{\exp\left(\frac{h\nu}{k_\mathrm B T}\right) - 1} = \frac{ 2 h \nu^3}{c^2} \frac 1{e^{h \beta \nu} - 1}.
\end{equation}
It is well known that the two agree for small values of $\nu$ since $e^x \simeq 1 + x + O(x^2)$ for $x\ll 1$.
\begin{equation}
    \frac{ 2 h \nu^3}{c^2} \frac 1{e^{h \beta \nu} - 1} = \frac{ 2 h \nu^3}{c^2} \frac 1{1 + h \beta \nu + O\left(\nu^2\right) - 1} \simeq \frac{2 \nu^2}{c^2 \beta}.
\end{equation}

Note that the same result can be achieved taking the limit for large temperatures, i.e.~the classical black-body radiation spectrum can also be understood as the first term in the expansion for small $\beta$,
\begin{equation}
    \frac{ 2 h \nu^3}{c^2} \frac 1{e^{h \beta \nu} - 1} = \frac{ 2 h \nu^3}{c^2} \frac 1{1 + h \beta \nu + O\left(\beta^2\right) - 1} \simeq \frac{2 \nu^2}{c^2 \beta}.
\end{equation}
Since $\beta$ is the inverse of the temperature, this is the limit for large temperatures. Since the entropy increases as the temperature increases, this is also the limit for large entropy. That is, the classical distribution is recovered as the high-entropy limit for the quantum distribution.

\subsection{Thermal equilibrium}

We can similarly reinterpret results that take the mathematical limit $\hbar \to 0$. For example, following the original paper from Wigner~\cite{WignerLimit}, we can consider the Wigner distribution of a system in thermal equilibrium at inverse temperature $\beta$:
\begin{equation}
    W(x,p)=\int dy e^{i(x+y)p/\hbar}\langle x+y |e^{-\beta\hat{H}}|x-y \rangle e^{-i(x-y)p/\hbar}
\end{equation}
Thermal equilibrium is described by $\hat{\rho}=e^{-\beta \hat{H}}$, which is the mixed state that maximizes entropy at a given average energy. The entropy of this state is directly connected to the value of $\beta$ such that when $S$ tends to $\infty$, $\beta$ tends to 0.
From this, Wigner considers the transformed Hamiltonian
\begin{equation}
    \tilde{H}=e^{i x p/\hbar}\hat{H} e^{-i x p/\hbar}=\frac{(p+i\hbar\partial/\partial x)^2}{2m}+V(x)=\epsilon(x,p)+i\frac{\hbar p}{m}\frac{\partial}{\partial x}-\frac{\hbar^2}{2m}\frac{\partial^2}{\partial x^2}
\end{equation}
where $\epsilon(x,p)$ is the classical Hamiltonian. The extra term contains the quantum corrections; we will refer to it as $\hat{Q}$. The Wigner function becomes
\begin{equation}
    W(x,p)=\int dy \langle x+y |e^{-\beta\tilde{H}}|x-y \rangle. 
\end{equation}
At this point, Wigner expands this expression in powers of $\hbar$, showing that quantum corrections are at second order. What we can do is instead expand in powers of $\beta$. We can show that quantum corrections are also only found at second order in $\beta$, justifying the classical limit. Let's start by considering the first-order expansion, $e^{-\beta\tilde{H}}\simeq 1-\beta(\epsilon+\hat{Q})$. So at first order
we get
\begin{equation}
    W(x,p)\simeq 1-\beta \epsilon(x,p)-\beta\int dy\langle x+y |\hat{Q}|x-y \rangle.
\end{equation}
Focusing on the last term, we can insert an identity in the momentum eigenbase
\begin{equation}
  \int dy\langle x+y |\hat{Q}|x-y \rangle=\int\int dkdy e^{i2ky} \left(-\frac{\hbar p}{m}k+\frac{\hbar^2}{2m}k^2\right)=\int dk \delta(2k) \left(-\frac{\hbar p}{m}k+\frac{\hbar^2}{2m}k^2\right)=0.
\end{equation}
The first-order quantum correction vanishes, showing that the classical approximation is correct to first order in $\beta$, which corresponds to the limit of large entropy. We can consider the second-order correction
\begin{equation}
     W(x,p)\simeq 1-\beta \epsilon(x,p)-\beta^2\epsilon^2(x,p)-\beta^2\int dy\langle x+y |\hat{Q}V(\hat{x})+V(\hat{x}) \hat{Q}|x-y \rangle,
\end{equation}
giving the first nonzero quantum correction to the Wigner function.

\section{The high-entropy limit}
\label{sec:high_entropy_limit}

Having given qualitative reasons for the limit, and having seen how, in specific cases, standard arguments can be reinterpreted consistently with our view, we now develop the limit as a direct consequence of entropy increase. Since the maximum entropy attainable by a quantum system is the logarithm of the dimension of the corresponding Hilbert space~\cite{nielsen2010quantum}, the high-entropy limit exists only for infinite-dimensional spaces. This explains why spin, which lives in a finite-dimensional space, is an intrinsically quantum property. Therefore we are going to concentrate on the case of a single degree of freedom identified by position and momentum.\footnote{A similar limit should exist for an infinite-dimensional directional degree of freedom (i.e.~for angular momentum). Though technically not a limit but more of an approximation, for finite-dimensional spaces one could look at mixed states that are close to the maximally mixed states (i.e.~$\rho = I/n + \epsilon \sigma$).} The case of multiple degrees of freedom can be recovered by increasing the entropy of all DOFs at the same time, that is, by increasing the entropy of each DOF independently, without introducing correlations.

The overall argument can be broken down into the following steps:
\begin{enumerate}
    \item Characterize entropy-increasing maps in classical mechanics for a single DOF, noting that they can all be understood as Hamiltonian evolution followed by a pure stretching map, one that stretches position and momentum by the same factor $\sqrt{\lambda}$. Under this map, the expectation of a polynomial of position and momentum of degree $n$ is also multiplied by a factor of $\sqrt{\lambda^n}$.
    \item Show that a pure stretching map in quantum mechanics cannot be defined over symmetrized operator averages or normal ordering.
    \item Show that it can be defined over anti-normal ordering. Show that if we rescale the zero of entropy, the commutators become $[X, P]= \frac{\imath \hbar}{\lambda}$, which makes the limit $\lambda \to \infty$ mathematically equivalent to $\hbar \to 0$.
    \item Show how, in the limit, the Wigner $W$ and Husimi $Q$ distributions become closer and closer.
\end{enumerate}

\subsection{Stretching classical phase space}

Before looking at the quantum case, let us study the high-entropy limit of classical mechanics. We are looking for all those transformations that increase the entropy of all states by the same amount.

We are going to study the one-dimensional case, therefore let us call $\mathcal{M} = (\mathbb{R}^2, \omega)$ the phase space for a single degree of freedom, where $\omega$ is the associated symplectic form. Suppose we have a map $R : \mathcal{M} \to \mathcal{M}$ that acts on phase space. This map will also act on probability distributions defined on $\mathcal{M}$ by moving them point to point. We require that $R$ increase the entropy of each distribution $\rho$ by a set value $\Delta S = \log \lambda > 0$. That is,
\begin{equation}
S(R(\rho)) = S(\rho) + \log \lambda,
\end{equation}
where $S(\rho) = - \int_{\mathcal{M}} \rho \log \rho dx dp$ and $R(\rho)$ is the distribution as it is transformed through the map.

We focus on maps that increase the entropy uniformly because, in the end, we will want Hamiltonian evolution on the original states to correspond to Hamiltonian evolution on the transformed states. Since Hamiltonian evolution maps states at constant entropy, $S(R(\rho))$ must be a function of $S(\rho)$. Moreover, note that for any $\rho_1$ and $\rho_2$ with disjoint support, $S\left(\frac{1}{2}\rho_1 + \frac{1}{2}\rho_2\right) = 1 + \frac{1}{2}S\left(\rho_1\right) + \frac{1}{2}S\left(\rho_2\right)$. Since $R$ will preserve disjointness of support, the one bit of entropy increase will have to carry through the map, meaning that the difference in entropy will need to be preserved. This is why we are focusing on maps that increase the entropy uniformly.

For a generic transformation, the increase of entropy is given by the expectation of the Jacobian determinant $\Delta S =\int_M \rho \log |\partial_a R^b| dxdp$. Since the increase has to be the same for all distributions, we must have $|\partial_a R^b| = \lambda > 1$.\footnote{The case of a negative Jacobian is a reflection in phase space, which cannot be achieved through a continuous evolution in time and is therefore discarded.} Recall that the Jacobian determinant tells us how an infinitesimal volume scales through the transformation, and therefore a map that increases entropy uniformly is exactly a map that stretches phase space uniformly. We call such a map a \textbf{stretching map}.

Suppose, in fact, that $\rho$ is a uniform distribution over a region of area $W_1$. Then we have $S(\rho)= \log W_1$. Applying the stretching map, $\rho$ will transform into a uniform distribution over a region of area $W_{\lambda} = \lambda W_1$. The final entropy is therefore $S(R(\rho)) = \log \lambda W_1 = \log W_1 + \log \lambda$. The factor $\lambda$, then, can be understood either as the ratio between initial and final areas, or as the exponential of the entropy increase. Therefore studying the high-entropy limit means studying what happens under stretching maps in the limit $\lambda \to \infty$.

Note that the transformation is not a canonical transformation. Canonical transformations, those that can be generated by Hamiltonian evolution, preserve areas in phase space and conserve entropy. In fact, for a single degree of freedom, canonical transformations and volume-preserving maps coincide. This allows us to show that all stretching maps can be written as $R = T \circ U$, where $U$ is a canonical transformation and $T$ is a \textbf{pure stretching map} defined as
\begin{equation}
    T(x,p) \mapsto (\sqrt{\lambda} x, \sqrt{\lambda} p)
\end{equation}
where $\lambda = (1,\infty)$. That is, any stretching map can be understood as first performing a canonical transformation that preserves entropy followed by a pure stretch of position and momentum.

To see how this works, note that given a stretching map $R$, we can always write $U = T^{-1} \circ R$. Since $R$ stretches phase space everywhere by a factor $\lambda$ and $T^{-1}$ shrinks it by the same factor, $U$ preserves areas and is a canonical map, since we restricted ourselves to the one-dimensional case.\footnote{To extend to the general case, the stretching map must not only stretch the total volume, but areas in each DOF. Mathematically, this means rescaling the symplectic form $\omega$.} Therefore $R = T \circ U$, which means we only need to study $T$ to characterize all maps that increase entropy uniformly.

We can alternatively characterize stretching maps by how the  Poisson brackets transform. Note that
\begin{equation}
    \{R(x),R(p)\} = \partial_x R(x) \partial_p R(p) - \partial_x R(p) \partial_p R(x) = | \partial_a R^b | = \lambda.
\end{equation}
That is, the Poisson bracket of the transformed position and momentum is the Jacobian determinant of the transformation, which is $\lambda$. Since this is an equality, all maps that satisfy the above transformation of the Poisson brackets are stretching maps.

For a pure stretching map, we can provide a more specific characterization. First of all, given that the map will act linearly on the space of probability distributions, once we characterize its action on distributions with compact support, we fully characterize the map. Furthermore, a distribution with compact support is always fully characterized by all its central moments, that is, the expectations for all polynomials of position and momentum (i.e.~the moment problem is always solvable). Therefore, once we know how the central moments transform through the map, we know how all distributions with compact support transform and therefore we fully characterize the map. This means that $T$ is a pure stretching map if and only if
\begin{equation}
	\langle T(x^np^m) \rangle = (\sqrt{\lambda})^{(n+m)} \langle x^n p^m \rangle.
\end{equation}
Given the linearity of the expectation, this relationship will remain true for all distributions that have finite moments. In particular, this will be true for all Schwartz functions. Therefore, we can use the above as the defining property of the pure stretching map.

\subsection{Operator ordering}

Ideally, we would want to treat the quantum case similarly to the classical case. We would look for those quantum channels (i.e.~CPTP maps) $R$ that increase the entropy of all mixed states by the same amount, we would show that they can be decomposed into a unitary part and a pure stretching part and so on. In practice, the quantum version of the problem is much harder. First of all, a map that increases the entropy of all quantum states cannot be invertible: pure states have the lowest entropy and therefore no state of lower entropy can be mapped to them. Secondly, since our map increases entropy, it does not in general preserve products. In fact, if we had $R(AB) = R(A)R(B)$, then $R([A,B]) = [R(A),R(B)]$, which would mean $R$ is unitary and therefore preserves entropy. Since, in general, $R(AB) \neq R(A)R(B)$, different permutations of polynomials will be transformed differently: the operator ordering matters.

In this light, we are looking for a map that rescales the expectations of the polynomials for a particular ordering. There are three commonly used orderings: symmetrized averages, normal ordering and anti-normal ordering. The first takes the product of $n$ operators by averaging all possible permutations; the second is in terms of polynomials of the form $(a^\dagger)^n a^m$, where $a=\sqrt{\frac{m\omega}{2\hbar}}(X+\frac{i}{m\omega}P)$ and $a^\dagger$ are the ladder operators, $m$ is the mass and $\omega$ the angular frequency; the last is in terms of $a^n (a^\dagger)^m$. The three orderings are associated, respectively, with a quasi-probability distribution~\cite{carmichael2013statistical}: the Wigner function $W$, the Glauber-Sudarshan $P$ distribution and the Husimi $Q$ distribution. In each case, the expectation under the quasi-probability distribution of polynomials of classical variables returns the expectation of the respective ordering.

Interestingly, not all orderings will allow a pure stretching map $T$ that rescales all expectations and increases the entropy of all states. Let us consider the symmetrized average case. We are looking for a map $T_W$ for which
\begin{equation}
\langle T_W(\Pi(\underbrace{X, ..., X}_{\text{n times}}, \underbrace{P, ..., P}_{\text{m times}})) \rangle = (\sqrt{\lambda})^{(n+m)} \langle \Pi(\underbrace{X, ..., X}_{\text{n times}}, \underbrace{P, ..., P}_{\text{m times}})\rangle,
\end{equation}
where $\Pi(A_1, A_2, \ldots, A_n)  = \frac{1}{n!} \sum_{\pi}  A_{\pi(1)} A_{\pi(2)} \cdots A_{\pi(n)}$ is the average of the products for each permutation $\pi$. These averages correspond to the expectation calculated through the Wigner function $W(x,p)$
\begin{equation}
     \langle \Pi(\underbrace{X, ..., X}_{\text{n times}}, \underbrace{P, ..., P}_{\text{m times}})\rangle = \int_M x^n p^m W(x, p) dx dp.
\end{equation}
The map $T_W$, then, would correspond to a pure stretching map on the Wigner function. However, this cannot work. Wigner functions can have regions with negative values, but the size of these regions cannot exceed a few units of $\hbar$~\cite{kenfack2004negativity}. The size of these negative regions would increase under $T_W$, giving functions that do not correspond to a quantum state. Therefore we cannot find a pure stretching map in the symmetrized average operator ordering.

Let us now consider a pure stretching map $T_P$ in normal ordering. This would have:
\begin{equation}
\langle T_P((a^\dagger)^n a^m) \rangle = (\sqrt{\lambda})^{(n+m)} \langle (a^\dagger)^n a^m \rangle.
\end{equation}
For the vacuum state we have $a|0\rangle=0$, which means the mean value of all observables in normal ordering for the vacuum is zero. These would remain unchanged by $T_P$. The vacuum state, then, would not change and therefore the map would not increase entropy for all states. The normal ordering is ruled out as well.

We turn our attention to the anti-normal ordering and a map $T_Q$ such that
\begin{equation}
\langle T_Q(a^n(a^\dagger)^m) \rangle = (\sqrt{\lambda})^{(n+m)} \langle a^n(a^\dagger)^m \rangle.
\end{equation}
This ordering solves the previous problem of the vacuum. The anti-normal ordering is connected to the Husimi $Q$ distribution by
\begin{equation}
    \langle a^n (a^\dagger)^m\rangle=\int \alpha^n(\alpha^*)^m Q_1(\alpha)d^2\alpha.
\end{equation}
The map $T_Q$, then, corresponds to a pure stretching map on the space where $Q$ is defined. That is,
\begin{equation}
T_Q(Q_1(\alpha)) \equiv Q_\lambda(\alpha) = \frac{1}{\lambda}Q_1\left(\frac{\alpha}{\sqrt{\lambda}}\right).
\end{equation}
We indicate $Q_1$ as the initial unstretched distribution and $Q_\lambda$ the final stretched distribution by a factor of $\lambda$. We can in fact verify that 
\begin{equation}
\int \alpha^n(\alpha^*)^m Q_\lambda(\alpha)d^2\alpha=\int \alpha^n(\alpha^*)^m \frac{1}{\lambda}Q_1\left(\frac{\alpha}{\sqrt{\lambda}}\right)d^2\alpha=\int \sqrt{\lambda}^{n+m}\beta^n(\beta^*)^m Q_1(\beta)d^2\beta.
\end{equation}
Since the Husimi distribution is non-negative, this avoids the issue presented by the Wigner function.

The anti-normal ordering, then, is a potential candidate. We now need to show that a pure stretching map $T_Q$ actually exists.

\subsection{Pure quantum stretching map}

If pure quantum stretching maps $T_Q$ exist, they must be CPTP maps as they must transform mixed states into mixed states. Moreover, since they have to increase entropy for all states, they must be describing an open quantum system. We thus show that pure stretching maps can be expressed as the solution of a master equation in Lindblad form and the Heisenberg picture. We start with
\begin{equation}
	\frac{d}{d t} X=\frac{i}{\hbar}[H,X]+\sum_i \gamma_i \left(L_i^\dagger X L_i-\frac{1}{2}\left\{L_i^\dagger L_i,X\right\}\right)
\end{equation}
where $X$ is an operator, $L_i$ are the jump operators and $\gamma_i$ are positive parameters. To define $T_Q$ for a single DOF we set $\gamma_i=\{\gamma\}$ and $L_i=\{a^\dagger\}$.\footnote{Since $a^{\dagger} |\psi\> = a^\dagger c_n |n\> = c_n \sqrt{n+1} |n+1\>$, $|\psi\>$ is in the domain of $a^\dagger$ if and only if $\sum |c_n|^2 n < \infty$ or, equivalently, the expectation $\<\psi|n|\psi\>$ is finite. Since every element of the basis $|n\>$ is in the domain, $a^\dagger$ is dense in $L^2$.} We also set $H=0$ since $T_Q$ should be purely dissipative. The equation of motion of $a$ reduces to
\begin{equation}
	\frac{d}{d t} a=\frac{\gamma}{2} a
\end{equation}
with solution $a(t)=e^{\frac{\gamma}{2} t} a = \sqrt{\lambda} a$, where we make the identification $\lambda=e^{\gamma t}$. 
In general, we find
\begin{equation}
	\frac{d}{d t} a^n (a^{\dagger})^m=\frac{\gamma}{2}(n+m)a^n (a^{\dagger})^m,
\end{equation}
meaning that this evolution realizes the stretching map in anti-normal ordering. That is, $(a^n (a^{\dagger})^m)(t)=e^{\frac{\gamma}{2} t (n+m)} a^n (a^{\dagger})^m = (\sqrt{\lambda})^{(n+m)} a^n (a^{\dagger})^m$.\footnote{Note that for the choice of $L_i=\{a\}$, we get the opposite behavior, characterized by
\begin{equation}
	\frac{d}{d t} (a^{\dagger})^n a^m=-\frac{\gamma}{2}(n+m)(a^{\dagger})^n a^m
\end{equation}
leading to shrinking ($\lambda\leq1$) of all observables taken in normal ordering.}

This shows that a map $T_Q$ that stretches the $Q$ distribution can be understood as a purely dissipative process that runs for a time $\Delta t$ with $\gamma_i = \left\{ \frac{\log \lambda} {\Delta t} \right\}$ and $L_i = \{ a^\dagger\}$, and therefore it is a CPTP map. Note that, for the same DOF, $a^\dagger$ is not uniquely fixed as it depends on the parameter $m\omega$. Moreover, ladder operators with different values of the parameter $m\omega$ are not going to commute, therefore will lead to different transformations. That is, while we fixed the ordering of the operators, the non-commutative nature of quantum observables still implies a choice within a one-parameter family of operators.  However, note that a linear transformation that stretches $X$ and shrinks $P$ as 
\begin{equation}
	U(X, P) = \left(\sqrt{\alpha} X, \frac{1}{\sqrt{\alpha}} P\right)
\end{equation}
is a unitary transformation. Under this map, $U(a) = \sqrt{\frac{m\omega\alpha}{2\hbar}}(X+\frac{i}{m\omega\alpha}P)$, which means that $m\omega$ can be changed through a unitary operator. Consistently with the classical definitions, we can define $R_Q = T_Q \circ U$ to be a quantum stretching map. Therefore, even if fixing the operator ordering does not pick a unique map, the behavior when entropy increases is the same.

Lastly, we need to show that $T_Q$ increases entropy for all states. Note that a Lindblad operator increases entropy for all states if~\cite{Benatti1988}
\begin{equation}
	\sum_i L_i L_i^\dagger \leq \sum_i L_i^\dagger L_i.
\end{equation}
In our case, this reduces to 
\begin{equation}
	a^\dagger a \leq a a^\dagger.
\end{equation}
Since $[a, a^\dagger] = a a^\dagger - a^\dagger a = I \geq 0$, our map increases entropy for all states.\footnote{Note that for the choice $L_i=\{a\}$ the inequality is not satisfied.}

We can now look at the effect of the map on $X$, $P$ and their commutator. We have
\begin{align}
    T_Q(a) &= \sqrt{\lambda} a \\ 
    T_Q(a^\dagger) &= \sqrt{\lambda} a^\dagger \\ 
    T_Q(X) &= T_Q\left(\sqrt{\frac{\hbar}{2m\omega}}(a^\dagger + a)\right) = \sqrt{\lambda} X \\
    T_Q(P) &= T_Q\left(\imath\sqrt{\frac{\hbar m \omega}{2}}(a^\dagger - a)\right) = \sqrt{\lambda} P \\
    T_Q([X, P]) &= T_Q(\imath \hbar) = \imath \hbar \\
    [T_Q(X), T_Q(P)] &= [ \sqrt{\lambda} X, \sqrt{\lambda} P] = \lambda [X, P] = \lambda \imath \hbar.
\end{align}
Recall that the uncertainty principle between two operators is linked to the commutator by the formula
\begin{equation}
    \sigma_A \sigma_B = \frac{1}{2} | \langle [A,B] \rangle |.
\end{equation}
Therefore we find that the uncertainty between $T_Q(X)$ and $T_Q(P)$ grows with $\lambda$, which is consistent with the stretching behavior of the map.

Note that when one takes the usual mathematical limit $\hbar \to 0$, the distribution, and therefore the uncertainty over position and momentum, is kept fixed. We can achieve that by redefining $X$ and $P$ in the following way
\begin{align}
    \hat{X} &= \frac{1}{\sqrt{\lambda}} X \\
    \hat{P} &= \frac{1}{\sqrt{\lambda}} P \\
    [\hat{X}, \hat{P}] &= \frac{\imath \hbar}{\lambda} \\
    T_Q(\hat{X}) &=  X \\
    T_Q(\hat{P}) &=  P \\
    [T_Q(\hat{X}), T_Q(\hat{P})] &= \imath \hbar.
\end{align}
Note how the redefinition of the operators effectively undoes the stretching operation. Also note that the redefinition reduces the minimum uncertainty between position and momentum. The same limit, then, can be understood in two different ways. In the more physical viewpoint, we are mapping states within the same state space, which means pure states remain at the same level of uncertainty, of entropy. The map will move states to higher entropy within the same space as $\lambda$ increases. Conversely, in the more mathematical viewpoint, we can redefine the state space while going through the map. The uncertainty, the entropy, of the mapped states will remain the same, but the pure states of the target space will have lower  uncertainty, lower entropy, effectively adding new states at these lower values. As we take the limit $\lambda \to \infty$ in the first viewpoint, in the second viewpoint the commutator $[\hat{X},\hat{P}] = \frac{\imath \hbar}{\lambda}$ decreases, which is equivalent to taking the limit $\hbar \to 0$. This is the group contraction that morphs the Moyal bracket Lie algebra to the Poisson bracket Lie algebra~\cite{Moyal_1949,saletan1961contraction, inonu1953contraction}. That is, as $\lambda$ increases, the entropy of pure states becomes lower and lower, and the structure of the quantum space becomes closer and closer to the structure of classical mechanics.

Note that while entropy could be increased in many different ways, under a stretching map unitary evolution will be mapped to another unitary evolution. In fact, if $H$ is a polynomial of position and momentum, we will be able to write the corresponding $\hat{H}$ such that $T_Q(\hat{H}) = H$. The stretching map, then, is useful to show not only that high-entropy states can be approximated by classical states, but that unitary evolution can be approximated by classical Hamiltonian mechanics. One may show that arbitrary entropy-increasing maps will recover classical states, for example by applying similar techniques to entropy bounds~\cite{Hall_2018}, but they will not necessarily recover classical Hamiltonian evolution.

\begin{figure}[h]
    \centering
\begin{tikzpicture}
\begin{axis}[
height=7cm,
width=\linewidth*0.8,
grid=both,
grid style={line width=.1pt, draw=gray!25},
axis lines=middle,
xmin=0,
xlabel = \(\Sigma\),
ylabel = \(S\),
legend style={at={(0.95,0.25)},anchor=east},
]
\addplot[thick,blue,samples=150,domain=0.06:1.5] {ln(x)+1};
\addlegendentry{\(\text{classical}\)}
\addplot[thick,BrickRed,densely dotted,samples=150,domain=0.5:1.5] {(x+1/2)*ln(x+1/2)-(x-1/2)*ln(x-1/2)};
\addlegendentry{\(\text{quantum}\)}
\addplot[thick,ForestGreen,dashed,samples=150,domain=0.25:1.5] {((2*x+1/2)*ln(2*x+1/2)-(2*x-1/2)*ln(2*x-1/2) - ln(2)};
\addlegendentry{\(\text{$\lambda = 2$}\)}
\addplot[thick,violet,densely dashed,samples=150,domain=0.05:1.5] {((5*x+1/2)*ln(5*x+1/2)-(5*x-1/2)*ln(5*x-1/2) - ln(5)};
\addlegendentry{\(\text{$\lambda = 5$}\)}
\end{axis}
\end{tikzpicture}
    \caption{Entropy $S$ in nats for a Gaussian state as a function of uncertainty $\Sigma$, measured in units of $\hbar$. It shows the effect of the transition from classical to quantum by shrinking the pure states by a factor of $\lambda$ instead of stretching the mixed states by the same factor. The rescaled uncertainty for pure states becomes $\hbar/\lambda$ and the modified Gaussian bound becomes $S_{\lambda Q}(\Sigma) = S_Q(\lambda \Sigma) - \ln(\lambda)$. Taking $\lambda$ to infinity corresponds to taking $\hbar/\lambda$ to zero.}
    \label{fig:uncertainty_scaled}
\end{figure}

As a more quantitative illustration, we can see what happens to the entropy of Gaussian states as we reduce the lowest possible uncertainty in the redefined quantum space. The entropy will become $S_{\lambda Q}(\Sigma) = S_Q(\lambda \Sigma) - \ln(\lambda)$. As we can see in Fig.~\ref{fig:uncertainty_scaled}, even with low factors of $\lambda$ the entropy curve approaches the classical one. If one takes the limit $\lambda \to \infty$, the above expression reduces to the classical equation. It should be stressed that these negative entropy states are fictitious, and it is only the difference in entropy that remains physically significant.

Note that the above argument works for any map $T_Q \circ U$ that combines the pure stretching map with an arbitrary unitary evolution $U$. Also, while we have not shown that anti-normal ordering is the only ordering that allows a stretching map, it will need to have the same effect on $X$ and $P$, leading to the same limit. The question of whether all maps that perform the same group contraction can be expressed as $T_Q \circ U$ remains open. However, they will factorize in that fashion in the limit, since the classical limit can always be factorized as a pure stretching map following a symplectomorphism. Similarly, the question of whether $T_Q$ increases the entropy of all states by the same amount remains open. However, it will do so in the limit since $T_Q$ will become closer and closer to a pure classical stretching map. Furthermore, we have not shown that all transformations that increase entropy must stretch phase space. Note, however, that a finite region of phase space can only hold mixed states with finite entropy. Therefore a map that does not stretch some region of phase space will necessarily lead to states whose entropy will not go to infinity as the map is reapplied over and over. Lastly, there is the question of whether any quantum channel that increases entropy will yield the classical limit. If by classical limit we mean the group contraction, then the only way to achieve it is through the rescaling of the commutator between $X$ and $P$, which is exactly a stretching map as we defined it. More in general, however, once a state has achieved sufficiently high entropy to be approximated by a classical state, it does not matter what process was used to achieve that high-entropy state. The remaining difficulty will be whether the evolution can still be approximated by a Hamiltonian evolution, which is likely to fail for a generic entropy-increasing process.\footnote{In the case that a process increases entropy differently in different regions of phase space, the evolution in the limit cannot be Hamiltonian, but may still be, for example, Newtonian.} Studying this more general case is outside of the scope of this work. Our goal is simply to show that, as entropy increases, the classical description becomes a suitable approximation. 

While we concentrated on the one-dimensional case, the result can be generalized to multiple dimensions by performing a similar stretching operation on all degrees of freedom simultaneously. That is, the entropy, as it increases, spreads equally not just within each degree of freedom, but also across degrees of freedom. This would give us $[T_Q(X^i),T_Q(P_i)] = \lambda \imath \hbar$, so that the same stretching factor is associated to all the appearances of $\hbar$ in the commutation relationship.

To sum up, we have found a pure quantum stretching map $T_Q$. It is a CPTP map. It rescales operators in anti-normal ordering. It increases entropy for all states. It recovers classical mechanics in the limit. This means that the space of quantum states, as entropy increases, becomes more and more classical. All of this is done under general conditions, and is independent of the mechanism that performs the entropy increase.

\subsection{Stretching the Wigner distribution}

To understand what happens physically during the group contraction, let us see how the Wigner $W$ distribution changes under a pure quantum stretching map. To do that, we can use the fact that $Q$ is the Weierstrass transform of $W$. That is,
\begin{equation}
    Q_\lambda(\alpha)=\frac{2}{\pi}\int W_\lambda(\beta)e^{-2|\alpha-\beta|^2}d^2\beta.
\end{equation}
Considering the evolution of $Q$, we get
\begin{equation}
    \int W_\lambda(\beta)e^{-2|\alpha-\beta|^2}d^2\beta=\frac{1}{\lambda}\int W_1(\beta)e^{-2|\alpha/\sqrt{\lambda}-\beta|^2}d^2\beta.
\end{equation}
We notice that these two equations are scaled convolutions between the Wigner distribution and Gaussian functions. Symbolically,
\begin{equation}
    (W_\lambda \ast G)(\alpha)=\frac{1}{\lambda}(W_1\ast G)\left(\frac{\alpha}{\sqrt{\lambda}}\right),
\end{equation}
where $G(\alpha)=(2/\pi) e^{-2|\alpha|^2}$ indicates the Gaussian function. To get rid of the convolution, we now take a Fourier transform of both sides. Keeping in mind that the Fourier transform of a general function $f$ behaves under scaling according to
\begin{equation}
    F\left(\frac{1}{\lambda}f\left(\frac{\alpha}{\sqrt{\lambda}}\right)\right)(k)=F(f(\alpha))(\sqrt{\lambda}k)
\end{equation} and using the convolution theorem, we get
\begin{equation}
  F(W_\lambda)(k)F(G)(k)=F\left(\frac{1}{\lambda}W_1\left(\frac{\beta}{\sqrt{\lambda}}\right)\right)(k)F(G)(\sqrt{\lambda}k).  
\end{equation}
We now notice that
\begin{equation}
    \frac{F(G)(\sqrt{\lambda}k)}{F(G)(k)}=\frac{e^{-\lambda|k|^2/8}}{e^{-|k|^2/8}}=e^{-(\lambda-1)|k|^2/8}=F(G)(\sqrt{\lambda-1}k)=F\left(\frac{1}{\lambda-1}G\left(\frac{\beta}{\sqrt{\lambda-1}}\right)\right)(k),
\end{equation}
finally giving
\begin{equation}
    W_\lambda(\beta)=F^{-1}\left(F\left(\frac{1}{\lambda}W_1\left(\frac{\beta}{\sqrt{\lambda}}\right)\right)F\left(\frac{1}{\lambda-1}G\left(\frac{\beta}{\sqrt{\lambda-1}}\right)\right)\right)=\frac{1}{\lambda(\lambda-1)}W_1\left(\frac{\beta}{\sqrt{\lambda}}\right) \ast G\left(\frac{\beta}{\sqrt{\lambda-1}}\right).
    \label{Fourier}
\end{equation}
Writing this down explicitly,
\begin{equation}
   W_\lambda(\beta)=\frac{2}{\pi\lambda(\lambda-1)}\int W_1\left(\frac{\alpha}{\sqrt{\lambda}}\right)e^{-\frac{2}{\lambda-1}\left|\alpha-\beta\right|^2}d^2\alpha.
\end{equation}
A similar calculation gives the Glauber-Sudarshan $P$ distribution, as this one is also obtained from a Weierstrass transfrom of $Q$.

It is interesting to consider what happens to the negative regions of $W$ under the stretching map. We know that $W$ can have negative regions, but their size is limited by the uncertainty principle. In fact, convolving $W$ with a 2D Gaussian with unitary spread, as in the definition of $Q$, returns a function that is never negative. In the limit $\lambda\gg1$, the formula for $W_\lambda$ reduces to
\begin{equation}
     W_\lambda(\beta)\rightarrow_{\lambda\gg1}\frac{2}{\pi\lambda^2}\int W_1\left(\frac{\alpha}{\sqrt{\lambda}}\right)e^{-\frac{2}{\lambda}\left|\alpha-\beta\right|^2}d^2\alpha=Q_\lambda(\beta).
\end{equation}
Therefore, while negative regions can be in principle found at any finite $\lambda$, $W$ tends to a positive function in the limit. As usual for the $W$ distribution, the phase-space size of negative regions is limited to $\hbar$ by the uncertainty principle. The weight of the function in such regions is limited between $\pm 2/\hbar$ for pure states. The effect of the stretching map is to reduce this bound to $\pm 2/(\lambda\hbar)$ for large values of $\lambda$. A clear interpretation can be made by working directly on the Fourier transform of $W$; see Eq.~(\ref{Fourier}). The function $F(W_\lambda)$ is a scaled version of $F(W_1)$ with a Gaussian filter applied to it. Quantum information in $W_1$ is carried by spectral weights with k-vectors larger than 1. The bandwidth of the Gaussian filter is given by $\lambda/(\lambda-1)$. This cutoff approaches 1 in the limit, filtering away the interference terms.

\begin{figure}[h]
	\centering
	\includegraphics[width=\textwidth]{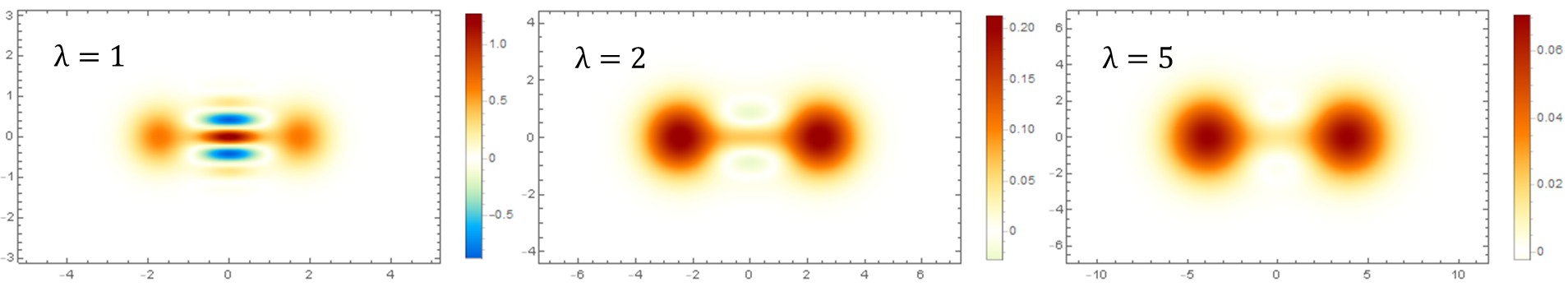}
	\caption{A cat state stretched with different values of $\lambda$. The axes are stretched by $\sqrt{\lambda}$ as well. The color scale is reduced to match the range of the function, though white always corresponds to zero. While the positive and negative peaks have comparable magnitude initially, the stretching map dampens the negative peak much more quickly. Note that, consistently with Fig.~\ref{fig:uncertainty_scaled}, the negative values are severely dampened with $\lambda=5$.}
	\label{fig:cat_streatched}
\end{figure}

In Fig.~\ref{fig:cat_streatched} we show a cat state as it is being stretched. Note that the stretching necessarily lowers the norm of the pseudo-probability density of both positive and negative regions. However, as expected, the negative regions are dampened much faster. After a stretching factor of $\lambda = 5$, the negative parts are negligible compared to the positive parts.

We have thus shown that, in the limit of high entropy, the Wigner distribution can be approximated by a positive function. We now need to show that the evolution can be approximated by classical Hamiltonian evolution. An intuitive way to understand why this works is to look at the evolution of the Wigner function under a Hamiltonian $H = p^2/2m + V$ where the potential $V$ is analytic. We have~\cite{hillery1984distribution}
\begin{equation}
\begin{aligned}
    \frac{d}{dt} W &= \{\{H, W\}\} \\
    &= \{H, W\} + \sum_n \frac{\hbar^{2n} (-1)^n}{(2n+1)! \, 2^{2n}} \partial_x^{(2n+1)} V \partial_p^{(2n+1)} W.
\end{aligned}
\end{equation}
The evolution is in terms of the Moyal bracket $\{\{H,W\}\}$, which can be expanded in orders of $\hbar$ with the Poisson bracket $\{H,W\}$ as the leading term. In our limit, $\hbar$ is constant but as the function stretches all derivatives of $W$ decrease. Therefore the first term becomes dominant.

It should be clear that this investigation does not exhaust all possible insights and links to the now vast literature on open quantum systems, quantum statistical mechanics, quantum information, quantum sensing and so on. For example, entropy and entropy increase play a role in several approaches: coarse-grained measurement (e.g.~Refs.~\cite{Kofler2007,bibak2025classicallimitquantummechanics}), decoherence (e.g.~Refs.~\cite{Zurek_RevModPhys.75.715,DecoherenceBook2003,SCHLOSSHAUER20191}), stochastic processes (e.g.~Refs.~\cite{carmichael2013statistical,QuantumStochasticReview,QuantumFokkerPlanck2022}) and information-theoretic and resource-theoretic approaches (e.g.~Refs.~\cite{Goold_2016,QuantumResourceReview2019,Shiraishi2025}) to name a few. On a more technical level, since both the Husimi Q and the Wigner function converge to the classical distribution, the Wehrl entropy~\cite{Wehrl1979}, which is the Shannon entropy calculated on the Husimi Q, will converge to the Gibbs entropy. One may ask, then, whether a high Wehrl entropy is already an indicator of classicality, and, if so, it would give another insight as to why the anti-normal ordering is important. Lastly, the role of entropic ideas is well established in historic literature. It would not be possible here to comment on all of them, and the more we mention, the more the omitted mentions are felt. Therefore we only explore the most important elements to establish the limit mathematically and under the smallest conceptual footprint, also as to not distract from the main point of the paper.

\section{Conclusion}

We have seen that classical mechanics can be recovered as the high-entropy limit of quantum mechanics. That is, states of higher and higher entropy are better and better approximated by classical distributions over phase space and, if the entropy increase is done uniformly, unitary evolution can be approximated with classical Hamiltonian evolution. This approach to the classical limit is independent of mechanism and interpretation, as it does not matter how the entropy is increased or what one believes quantum states to represent: as long as the description is in terms of mixed states of sufficiently high entropy, classical mechanics applies. The approach fits naturally with experimental considerations and other approaches, such as decoherence, leading to an interpretation that is  thermodynamically meaningful, representation-independent, and stable under coarse-graining. It also recovers the established mathematical recipe, which is the group contraction for $\hbar \to 0$.\footnote{As pointed out by a reviewer, the $\hbar \to 0$ limit can be problematic for chaotic or non-analytic systems. Therefore there may be an additional concrete advantage to our approach over the standard one.} Physically, this can be understood as taking the entropy of pure states to minus infinity, which is equivalent to saying that the relative entropy of mixed states goes to plus infinity. That is, in the same way that $c \to \infty$ should be understood as $v \ll c$, $\hbar \to 0$ should be understood as $S \gg 0$, as physically we cannot take limits of physical constants. This gives a reasonable and precise account of the classical limit: in the same way that non-relativistic mechanics applies to low speed, classical mechanics applies to high entropy.

The limit also gives us additional insights. As we saw, classical states for which $\sigma_x \sigma_p < \frac{\hbar}{e}$ are states with negative entropy, which imply a breakdown of thermodynamics. It is no wonder, then, electrons cannot fall on the nucleus or that classical mechanics gives us the wrong spectra for black-body radiation. Classical mechanics fails at low entropy. The use of the correspondence principle as guidance for the development of quantum mechanics, then, can be understood as requiring that the new quantized theory reproduce classical mechanics at high entropy. 

In this light, we wonder whether it can be proven that quantum mechanics is the only way to fix the low-entropy range of classical mechanics. That is, is quantum mechanics the only theory that can recover classical mechanics at high entropy? To us, this is a question whose answer would be interesting regardless of the outcome.

\section*{Acknowledgements}
This paper is part of the research program Assumptions of Physics~\cite{aop-book}, which aims to identify a handful of physical principles from which the basic laws can be rigorously derived.

\paragraph{Funding information}
ML acknowledges funding via the FWF-funded QuantERA project QuSiED with project number I 6008-N.

\bibliographystyle{plain}
\bibliography{bibliography}

\end{document}